\documentclass[a4paper,11pt]{article}

\usepackage{rotating}
\usepackage{graphics}
\usepackage{latexsym}
\usepackage[dvips]{epsfig}
\usepackage{amssymb}
\usepackage{graphicx}
\usepackage[ruled,linesnumbered,vlined]{algorithm2e}

\usepackage{url}

\def\boxit#1{\vbox{\hrule\hbox{\vrule\kern4pt
  \vbox{\kern1pt#1\kern1pt}
\kern2pt\vrule}\hrule}}

\def\qed{\rule{1.5mm}{3mm}}

\newcommand\nc{\newcommand}

\newtheorem{theorem}{\bfseries Theorem}
\newtheorem{lemma}[theorem]{Lemma}

\newtheorem{defn}[theorem]{Definition}

\newtheorem{example}{Example}
\newtheorem{corollary}{Corollary}
\newtheorem{definition}{Definition}

\nc{\crl}[2]{\begin{corollary}\label{crl:#1} #2 \end{corollary}}
\nc{\dfn}[2]{\begin{defn}\label{def:#1} #2 \end{defn}}
\nc{\lem}[2]{\begin{lemma}\label{lem:#1} #2 \end{lemma}}
\nc{\prp}[2]{\begin{proposition}\label{prp:#1} #2
\end{proposition}}
\nc{\thm}[2]{\begin{theorem}\label{thm:#1} #2\end{theorem}}
\nc{\fac}[2]{\begin{lemma}\label{fact:#1} #2 \end{lemma}}

\nc{\eqn}[2]{\begin{eqnarray}\label{eqn:#1} #2 \end{eqnarray}}

\nc{\fig}[4]{\begin{figure}[h]
\begin{center}
\includegraphics[width=#2\textwidth]{#4}
\end{center}
\caption{#3}\label{fig:#1}
\end{figure}}

\nc{\tbl}[3]{\begin{table}[hbt] #3 \caption{#2} \label{tab:#1}
\end{table}}

\nc{\refc}[1]{Corollary~\ref{crl:#1}}
\nc{\refd}[1]{Definition~\ref{def:#1}}
\nc{\reff}[1]{Fig.~\ref{fig:#1}}
\nc{\refl}[1]{Lemma~\ref{lem:#1}}
\nc{\refp}[1]{Proposition~\ref{prp:#1}}
\nc{\reft}[1]{Theorem~\ref{thm:#1}} \nc{\refe}[1]{(\ref{eqn:#1})}
\nc{\reftb}[1]{Table~\ref{tab:#1}}

\nc{\reffc}[1]{Fact~\ref{fact:#1}}

\nc{\pf}[1]{ \noindent \emph{Proof.} #1
 \hfill \qed\par}

\renewcommand{\title}[1]{\vspace{\fill}
\eject\addtolength{\baselineskip}{4pt}
{\bfseries\Large #1}\\[3mm]\addtolength{\baselineskip}{-4pt}}
\renewcommand{\author}[3]{\parbox[t]{75mm}
{\begin{center}{\scshape #1}\\[3mm] #2\\
 {\ttfamily #3} \end{center}}}

\long\def\invis#1{}

\begin{document}

\begin{center}
\title{Algorithms for Manipulating Sequential Allocation}

\author{Mingyu Xiao}
 {School of Computer Science and Engineering,
University of Electronic Science and Technology of China, China,
}{myxiao@gmail.com}
\author{Jiaxing Ling}{
School of Computer Science and Engineering,
University of Electronic Science and Technology of China, China,
}{251093927@qq.com}
\end{center}

\begin{abstract}
Sequential allocation is a simple and widely studied mechanism to allocate indivisible items in turns to agents according to a pre-specified picking sequence of agents. At each turn, the current agent in the picking sequence picks its most preferred item among all items having not been allocated yet. This problem is well-known to be not strategyproof, i.e., an agent may get more utility by reporting an untruthful preference ranking of items. It arises the problem: how to find the best response of an agent?
 It is known that this problem is polynomially solvable for only two agents and NP-complete for arbitrary number of agents.
 The computational complexity of this problem with three agents was left as an open problem. In this paper, we give a novel algorithm that solves the problem in polynomial time for each fixed number of agents. We also show that an agent can always get at least half of its optimal utility by simply using its truthful preference as the response.
%
\end{abstract}

\section{Introduction}
\par

Sequential allocation is a simple and widely studied mechanism to allocate indivisible items to agents~\cite{DBLP:conf/ijcai/BouveretL11,DBLP:conf/ecai/BouveretL14,DBLP:conf/aaai/AzizBLM17}.
In a sequential allocation mechanism, there are several indivisible items to be allocated to some agents,
each agent has a strict preference ranking over all the items, and there is a sequence of the agents, called the \emph{policy}, to specify the turns of the agents to get the items.
The items are allocated to the agents according to the policy:
at each turn, the current agent on the policy picks the
most preferred item in its preference ranking that has not yet been allocated.
We give an example.

\begin{example}\label{ex1} There are five items $\{a,b,c,d,e\}$, three agents $\{1,2,3\}$ with preference rankings
$$\mbox{Agent 1}: a\succ b \succ c\succ d\succ e$$
$$\mbox{Agent 2}: c\succ b \succ e\succ d\succ a$$
$$\mbox{Agent 3}: e\succ b \succ d\succ c\succ a$$
and a policy
$$\pi:13221.$$
\end{example}
In this example, Agent 1 will take $a$ at the first turn, Agent~3 will take $e$ at the second turn, Agent 2 will take two items $c$ and $b$ at the third and fourth turns, and Agent 1 will take $d$ at the last turn.

In sequential allocation, given a fixed policy, the outcome will only depend on the ordinal preference rankings of agents over items.
It is folklore that sequential allocation is not strategyproof, which means that an agent may get more utility by reporting an untruthful preference ranking.
For example, in the above instance, if Agent 1 misreports its preference ranking as $ b\succ a \succ c\succ d\succ e$, then it will get items $\{b,a\}$, while
originally it will get items $\{a, d\}$. Agent 1 may get more utility by taking $\{b,a\}$ since $b\succ d$.
This motivates many aspects of study on this mechanism.

There are several models based on the sequential allocation mechanism. We have different objectives to maximize the overall social welfare~\cite{DBLP:conf/ijcai/BouveretL11} or the utility of a certain agent~\cite{DBLP:conf/ecai/BouveretL14}, and different requirements on the pattern of the picking sequences and the number of agents.
One of the earliest models studied in~\cite{DBLP:journals/ior/KohlerC71} has two agents and the policy is strictly alternating $(e.g., 121212\dots)$.
A balanced alternation pattern of the policy $(e.g., 12212112\dots)$ was studied in~\cite{DBLP:books/daglib/0017729}.
One interesting application of sequential allocation was found in course allocation to students and several axiomatic properties and manipulability on this application have been revealed.
Budish and Cantillion~\cite{budish2012multi} investigated a randomized version of the sequential allocation mechanism
to allocate courses to students, and Pareto optimal solutions for a model of course allocation were studied in~\cite{cechlarova2018pareto}.
The Boston mechanism is another sequential allocation mechanism with applications in school choice for students~\cite{abdulkadiroglu2006changing,kojima2010boston}.
A general and systematic study of the sequential allocation was done by Bouveret and Lang~\cite{DBLP:conf/ijcai/BouveretL11}
and by Kalinowski et al.~\cite{DBLP:conf/ijcai/KalinowskiNW13} from a game-theoretic view.
Since the work of Kohler and Chandrasekaran~\cite{DBLP:journals/ior/KohlerC71}, a series of followup works on strategic aspects of sequential allocation
have been made~\cite{DBLP:conf/aldt/0001GW17,DBLP:conf/ijcai/AzizWX15,DBLP:journals/tamm/LevineS12,DBLP:conf/atal/TominagaTY16,DBLP:conf/atal/AzizGMMNW15}.

In this paper, we consider manipulations in sequential allocation.
In this model, the policy is given, and among all agents, one is the \emph{manipulator} and all others are \emph{non-manipulators}.
The manipulator needs to report a list of items as its preference ranking
to achieve a certain objective.
There are two commonly used assumptions. Firstly,
the manipulator has complete information about the reported preferences
of non-manipulators.
This is a worst case assumption often made in computer science and
economics.
Secondly, the manipulator has additive cardinal utilities for the
items, although agents report strict and ordinal preferences.
This assumption is standard in this research area.

We can define several problems with different objectives of the problem.
The \textsc{Best Response} problem is to find a best response of the manipulator (i.e., a preference ranking which allows it to obtain
the maximum utility). \textsc{Better Than Truth Response} is to ask whether the manipulator can get more utility than
the allocation under its truthful report.
\textsc{Allocation Response} is to ask whether the manipulator can get a specified bundle of items.
Among all these problems, \textsc{Best Response} seems to be the hardest one and a solution to it can imply solutions to other problems,
since other problems can be easily reduced to \textsc{Best Response}.
See~\cite{azizmanipulating} for a recent survey
on the results for these problems.

For \textsc{Best Response},
Bouveret and Lang~\cite{DBLP:conf/ijcai/BouveretL11} first showed that the problem with only two agents (one manipulator and one non-manipulator) can be solved
in polynomial time. Then Aziz \emph{et al.}~\cite{DBLP:conf/aaai/AzizBLM17} proved that it is NP-hard to compute the best response of the manipulator if the number of agents is part of the input by correcting a wrong claim in a previous paper. It becomes an open problem whether \textsc{Best Response} is polynomially solvable for three or another constant number of agents~\cite{DBLP:conf/aaai/AzizBLM17}.
This open problem is interesting because it is already known that the problem is polynomially solvable with the utility functions of the manipulator being some specified functions, such as lexi-cographic utilities and binary utilities~\cite{DBLP:conf/ijcai/BouveretL11,DBLP:conf/aaai/AzizBLM17}.
In this paper, we fully answer this question by giving a dynamic programming algorithm for \textsc{Best Response} that runs in polynomial time for any fixed number of agents and any additive utility functions.
In addition, we show that  the manipulator can always get at least half of the optimal utility if it simply uses the truthful preference ranking, where the approximation
ratio is tight as far as using the truthful preference ranking.


\section{Preliminaries}
In the  sequential allocation problem, $m$ items are going to be allocated to $n$ agents
according to a \emph{policy} $\pi$, which is a sequence of agents specifying the turns of the agents to get items.
The length $|\pi|$ of the policy is $m$ since there are $m$ items to be allocated.
The set of items is denoted by $O=\{g_1,g_2,\dots, g_m\}$ and the set of agents is denoted by $N=\{1,2,\dots,n\}$, where Agent 1 is the manipulator and all other agents are non-manipulators.  Each agent $i\in N$ has a complete preference ranking $\succ_i: g_{i_1},g_{i_2},\dots, g_{i_m}$ over all items in $O$.
We will write $g_{p} \succ_i g_{q}$ to denote that item $g_{p}$ is ranked ahead
of $g_{q}$ in Agent~$i$'s preference ranking.
The manipulator (Agent 1) has an additive utility function on the items $u: O \rightarrow \Re+$.
For two items $g_x, g_y \in O$, it holds $u(g_x)> u(g_y)$ if and only if $g_{x} \succ_1 g_{y}$.
We use $k_i$ ($i\in N$) to denote the \emph{frequency} of Agent~$i$ appearing in the policy $\pi$, and use $m'$ to denote
the frequency of non-manipulators appearing in $\pi$. Then it holds that
$$ m=\sum_{i=1}^n k_i, ~~\mbox{and}~~  m'=m-k_1.$$

For \textsc{Best Response}, the manipulator wants to find a \emph{picking strategy} to achieve its
maximum utility, i.e., a permutation of all the items, according to which to pick up items the manipulator can get the maximum utility.
When we say a \emph{solution} to \textsc{Best Response}, it is regarded as the optimal picking strategy or the bundle of items for the manipulator determined by the optimal picking strategy.
We use $I=(O, N, \pi, \{\succ_i\}_{i=1}^n)$ to denote our input instance, where we omit the utility function of the manipulator
to simplify the description since for most cases we only use the preference ranking $\succ_1$.

Once a picking strategy is given, we will get a fixed sequence of allocations of all items to agents, called \emph{allocation sequence}.
We will say the above picking strategy and allocation sequence are \emph{associated} with each other.
If there is no picking strategy associated with an allocation sequence, then the allocation sequence is called \emph{infeasible}; otherwise, it is called \emph{feasible}.
For a feasible allocation sequence, it is easy to construct one picking strategy associated with it.

A \emph{partial allocation sequence} is a sub sequence of an allocation sequence beginning from the first allocation.
We will use $\xi$ to denote an allocation sequence and use $\xi(i)$ to denote the partial allocation sequence of the first $i$ allocations of $\xi$.
For each \emph{feasible} partial allocation sequence of length $l$, there is a \emph{partial} policy of length $l$
and a \emph{partial} picking strategy associated with it.
After executing a partial allocation sequence according to a partial policy, we will get a \emph{remaining problem} which is to allocate the remaining items to the agents according to the remaining policy.

Given a (partial) allocation sequence, we say an item $g$ has been \emph{considered} by Agent~$i$ before the $x$th position of the (partial) policy if during the first $x$ allocations in the sequence the last item allocated to Agent~$i$ is ranked lower than item $g$ in  Agent~$i$'s preference ranking. Note that an item may not be allocated to an agent even if the item has been considered by the agent.

\begin{figure}[ht]
    \centering
    \includegraphics[width = 1\linewidth]{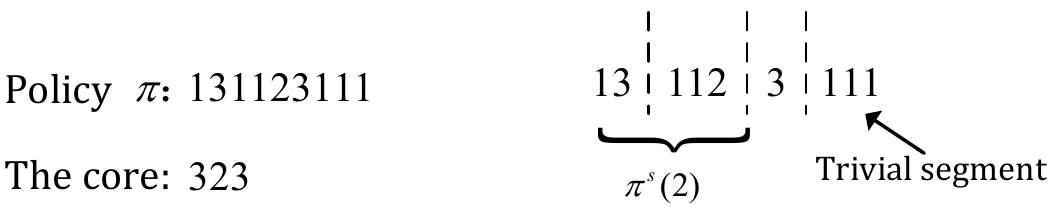}
    \caption{The segments and core}\label{fig1}
\end{figure}

A \emph{segment} in a policy is a maximal continuous subsequence containing at most one position of a non-manipulator and only the last position of the subsequence can be the non-manipulator.
 A policy having $m'$ positions of non-manipulators can be partitioned into $m'+1$ segments by cutting after each non-manipulator position, where the last segment is called a \emph{trivial segment}. A trivial segment only contains copies of the manipulator and it may be empty (when the last position of the policy is a non-manipulator). A nontrivial segment may contain only one non-manipulator. We will use $\pi^s(x)$ to denote the partial policy of the first $x$ segments of $\pi$.
 The \emph{core} of a (partial) policy is the sequence of agents obtained by deleting
 all occurrences of the manipulator from the (partial) policy. See Figure~\ref{fig1} for an illustration of the segments and cores.

The \emph{position vector} of the manipulator in a (partial) policy $\pi$ is a sequence of increasing positive integers,
$(z_{1},z_2,\dots, z_{k_1})$ to denote the positions of the manipulator in the policy $\pi$, i.e., the manipulator appears on the $z_{1}$th, $z_2$th, $\dots$, and $z_{k_1}$th positions in the (partial) policy $\pi$.
A  policy $\pi$ \emph{dominates} another policy $\pi'$ if they have the same length and the same core and
it holds that $z_i\leq z'_i$, $i\in \{1,2,\dots,k_1\}$, for manipulator position vectors in
$\pi$ and $\pi'$ being $(z_{1},z_2,\dots, z_{k_1})$ and $(z'_{1},z'_2,\dots, z'_{k_1})$, respectively.
This is to say that $\pi'$ can be obtained from $\pi$ by iteratively moving a manipulator in it to the next position.
For two instances $I=(O, N, \pi, \{\succ_i\}_{i=1}^n)$ and $I'=(O, N, \pi', \{\succ_i\}_{i=1}^n)$
with only different policies, if $\pi$ dominates $\pi'$, then we say instance $I$ \emph{dominates} instance $I'$.

Our algorithm for \textsc{Best Response} uses two major ideas.
The first idea is to reduce instances to constrained instances, called ``crucial instances''.
Crucial instances can be solved quickly directly. However, it is not easy to find the corresponding crucial instances and we still need to search among a large number of candidates.
So we also use the second idea, which is a divide-and-conquer technique, to reduce the number of candidates.
 The divide-and-conquer method will split the allocation problem into two subproblems:
the first one is to allocate a fixed set of items and the second one is to allocate the remaining set of items.
To guarantee that we can combine optimal solutions to the two parts to construct an optimal solution for the whole problem, we need some ``invariance properties''. Based on ``invariance properties'', we are able to design a dynamic programming algorithm to save running time.
We first introduce the two ideas in the following two sections.

\section{Crucial Instances}
In \textsc{Best Response},
we may have the same optimal picking strategy for two instances with only different policies.
These instances have some common properties.
We will classify some instances (and their policies) that have the same optimal picking strategy and solution into a class.
In each class, there is a special instance, called ``crucial instance'', which can be solved directly.
So we will try to solve an instance by solving the corresponding crucial instance in the same class.
This is the rough idea of our algorithm.

We give an example to illustrate that two instances with only different policies have the same optimal solution.
In Example~\ref{ex1}, the manipulator gets the best bundle $\{a,b\}$ by using picking strategy
$bacde$. We use $I'$ to denote the instance after replacing the policy $\pi:13221$ with policy $\pi':32121$.
In $I'$, the manipulator can get the same best bundle $\{a,b\}$ by using the same picking strategy.
Compared with $\pi$, the manipulator has a lower priority to pick items in $\pi'$.
However, the manipulator still can get the best solution.
Note that,  at the first position of the policy $\pi$, the manipulator picks an item that will not be considered by any non-manipulator before the $3$rd allocation.
So we can delay the allocation of $b$ to Agent 1 from position 1 to position 3 without changing the optimality.
Given an instance, we want to know how much we can delay the positions of the manipulator without losing the optimality and the ``worst'' policy
will be ``crucial''.

\vspace{2mm}
\begin{definition}[\textbf{Crucial Instance}]
For an instance $I=(O, N, \pi, \{\succ_i\}_{i=1}^n)$, if for any policy $\pi'\neq \pi$ dominated by $\pi$, the optimal solution to  the dominated instance $I'=(O, N, \pi', \{\succ_i\}_{i=1}^n)$ is worse than that to $I$, then we say $I$ is a \emph{crucial} instance.
Let $I'=(O, N, \pi', \{\succ_i\}_{i=1}^n)$ be a crucial instance, for any instance $I=(O, N, \pi, \{\succ_i\}_{i=1}^n)$
dominating $I'$, we say $I'$ is a \emph{corresponding} crucial instance to $I$.
\end{definition}

A corresponding crucial instance of an instance may be itself when it is already a crucial instance.
To solve an instance, we can turn to solve a corresponding crucial instance by the following lemmas.

\begin{lemma}\label{lem1}
 Given two instances $I=(O, N, \pi, \{\succ_i\}_{i=1}^n)$ and $I'=(O, N, \pi', \{\succ_i\}_{i=1}^n)$,
  where $I$ dominates $I'$. By using the same picking strategy, the manipulator in instance $I$ will get a bundle with total utility
  not less than that in $I'$. Furthermore, for any picking strategy $S'$, there is a picking strategy $S$ such that
   by using $S$ in $I$ the manipulator can get the same bundle as that by using $S'$ in $I'$.
\end{lemma}

\pf{
The first claim is easy to observe. We focus on the second claim.

We define the picking strategy $S$ to $I$ as follows: order the items according to the ordering of allocations to the manipulator in $I'$ by using $S'$, i.e., an item is ranked on the $i$th position in $S$ if it is the $i$th item allocated to the manipulator in $I'$, and all other items not allocated to the manipulator in $I'$ are listed behind with any order. Let $(z_{1},z_2,\dots, z_{k_1})$ and $(z'_{1},z'_2,\dots, z'_{k_1})$
be the position vectors of the manipulator of $\pi$ and $\pi'$. Since $\pi$ dominates $\pi'$,
we know that $z_i\leq z'_i$ for any $i\in \{1,2,\dots,k_1\}$. If an item can be allocated to the manipulator at position $z'_i$ in $\pi'$, then it can also be allocated to the manipulator at position $z_i\leq z'_i$ in $\pi$, since only a subset of items have been allocated before position $z_i$ in $\pi$ (compared to the situation at position $z'_i$ in $\pi'$).
So at each position, the manipulator can always get the current item on its picking strategy $S$. By using $S$,
the manipulator in $I$ gets the same bundle as that by using $S'$ in $I'$.
}

\begin{lemma}
Let $I=(O, N, \pi, \{\succ_i\}_{i=1}^n)$ be an instance and $S$ be a picking strategy for the manipulator.
Assume that by taking $S$ the manipulator picks an item at the $i$th position of $\pi$ and this item is not considered by any non-manipulator before the $j$th allocation, where $j>i+1$. Let $\pi'$ be the new policy obtained from $\pi$ by moving the manipulator from the $i$th position to the $(j-1)$th position.
 By using picking strategy $S$ in $I'=(O, N, \pi', \{\succ_i\}_{i=1}^n)$ the manipulator can get the same bundle
 as that by using $S$ in $I$.
\end{lemma}


\pf{


We consider the allocation sequence under the picking strategy $S$ in $I$:
$g_{1} \mapsto_1 l_1, g_{2} \mapsto_2 l_2, \dots, g_{m} \mapsto_m l_m$, where $l_1, l_2,\dots,l_m \in N$ and it is possible $l_j=l_k$ for $j\neq k$.
In the allocation sequence, item $g_i$ is allocated to the manipulator at position $i$, and before allocating the $j$th good $g_j$, the item $g_i$ has never been considered by any non-manipulator. Therefore, moving the $i$th position to the $j$th position in $\pi$ will not affect the allocations on the $(i+1)$th to $(j-1)$th positions. Furthermore,
after allocating these items, we still can allocate item $g_i$ to the manipulator on the $j$th position since it is still available. All other allocations will keep unchanged.
Thus, the following is still a feasible allocation sequence for $I'$: $g_{1} \mapsto_1 l_1, \dots, g_{i-1} \mapsto_{i-1} l_{i-1}, g_{i+1} \mapsto_{i} l_{i+1}, \dots, g_{i} \mapsto_{j-1} l_{i}, g_{j} \mapsto_{j} l_{j}, \dots, g_{m} \mapsto_m i_m$, in which the manipulator will get the same bundle as that in $I$.
}

\begin{corollary} \label{cor1}
Let $I=(O, N, \pi, \{\succ_i\}_{i=1}^n)$ be an instance and $I'=(O, N, \pi', \{\succ_i\}_{i=1}^n)$ be a corresponding  crucial instance. An optimal picking strategy for $I'$ is also an optimal picking strategy for $I$.
\end{corollary}

Next, we show how to solve crucial instances.

\begin{lemma}\label{lem_crucial}
Crucial instances of \textsc{Best Response} can be solved in linear time.
\end{lemma}

To prove Lemma~\ref{lem_crucial}, we use an algorithm to solve crucial instances optimally.
The algorithm is a greedy algorithm, called GreedyAlg.
We introduce the algorithm separately below because it will also be used in several places later.

\medskip
\textbf{Algorithm GreedyAlg}
The algorithm GreedyAlg takes a (sub) instance of \textsc{Best Response} as the input,
and outputs an allocation sequence with the corresponding picking strategy for the manipulator. However, the output allocation sequence may not be optimal for non-crucial instances.

The main idea of the algorithm is as follows. We allocate items to agents according to the policy.
Assume that we have allocated the first $i-1$ items to agents in the input instance $I=(O, N, \pi, \{\succ_i\}_{i=1}^n)$.
If the $i$th position in the policy $\pi$ is a non-manipulator, we let the non-manipulator pick its most preferred item that has not yet been allocated. Next, we consider the situation that the $i$th position in the policy $\pi$ is the manipulator.
The algorithm decides the item that should be assigned to the manipulator at the $i$th turn by the following method.

We let $I'=(O', N, \pi', \{\succ'_i\}_{i=1}^n)$ be the remaining instance after allocating the first $i-1$ items in $I$.
Then the first position in $\pi'$ is the manipulator.
Let $\pi_{ -1}$ be the core of $\pi'$.
If $\pi_{ -1}$ is empty, we assign the best remaining item $o$ in the truthful preference ranking of the manipulator
to the manipulator at the $i$th position of $\pi$ and let $o$ be the $i$th object in the picking strategy of the manipulator.
If $\pi_{ -1}$ is not empty, we let $o_f$ be
the favourite item in $O'$ of the first agent in $\pi_{-1}$.
GreedyAlg will assign item $o_f$ to the manipulator at the $i$th position of $\pi$ and let $o_f$ be the $i$th object in the picking strategy of the manipulator.

According to the above method, the algorithm decides the items to be assigned to the manipulator from the first
occurrence of 1 to the last occurrence of 1 in $\pi$ and then we can get a full allocation sequence and a picking strategy for the manipulator. This is the algorithm GreedyAlg.

It is easy to see that GreedyAlg can be implemented in linear time and the picking strategy returned by GreedyAlg for each instance is unique.

The allocation sequence returned by GreedyAlg on a (sub) instance is called \emph{greedy}.
Given an allocation sequence, we can easily check whether it is greedy or not.
The concept of greedy allocation sequence is also important and will be used later.

The correctness of Lemma~\ref{lem_crucial} directly follows from the following lemma.

\begin{lemma}\label{lem_greedy}
The greedy strategy for a crucial instance is the optimal solution to it.
\end{lemma}
In fact, a crucial instance has only one optimal allocation
sequence, which is the greedy one obtained by GreedyAlg.
Note that if in a solution, the manipulator at the $i$th position of $\pi$ does not pick
item $o_f$, then we could move the $i$th position of 1 to the behind of the first non-manipulator in $\pi$ to get a dominated instance $I'$, where $I'$ has the same optimal solution as $I'$, which contradicts the fact that $I$ is a crucial instance.

Although crucial instances can be solved quickly, it is still hard to find a corresponding crucial instance for an arbitrary instance.
We need to reveal more properties for dominated instances.

Lemma~\ref{lem1} implies that the optimal solution to an instance is not worse than the optimal solution to any dominated instance.
Clearly, the opposite direction of Lemma~\ref{lem1} may not hold.
We prove the following lemma.

\begin{lemma}\label{search}
Let $I$ be an instance and $\mathrm{P}$ be the set of instances dominated by $I$.
For each $I'\in \mathrm{P}$, we use $G(I')$ to denote the greedy allocation sequence of $I'$.
Assume that the greedy allocation sequence $G(I_0)$ $(I_0\in \mathrm{P})$ gets the best solution among all $G(I')$ with $I'\in \mathrm{P}$.
Then $I_0$ is a crucial instance corresponding to $I$.
\end{lemma}

The correctness of this lemma follows from Lemma~\ref{lem1}, Corollary~\ref{cor1} and Lemma~\ref{lem_greedy}.
Lemma~\ref{lem1} says any dominated instance $I'$ will not have a better solution than $I$.
Corollary~\ref{cor1} says that there is at least one dominated instance, the corresponding crucial instance will achieve the same optimal solution to $I$.
The greedy allocation sequence may not be optimal for any instance.
But it is optimal for a crucial instance by Lemma~\ref{lem_greedy}.
Therefore, among all the greedy allocation sequences, the best one is for a corresponding crucial instance.

\medskip
Lemma~\ref{search} implies that we can solve \textsc{Best Response} by taking each dominated instance as a candidate for a corresponding crucial instance and use GreedyAlg to solve it. We analyze the running time of this algorithm.
Let $(z_{1},z_2,\dots, z_{k_1})$ and $(z'_{1},z'_2,\dots, z'_{k_1})$
be the position vectors of the manipulator of two policies $\pi$ and $\pi'$.
We know that  $\pi$ dominates $\pi'$ if and only if  $z_i\leq z'_i$ holds for any $i\in \{1,2,\dots,k_1\}$.
The length of these policies is $m$.
So for $i\in \{1,2,\dots,k_1\}$, the value of $z'_i$ can be any integer between $\max\{z_i, z'_{i-1}+1\}$ and $m$.
Combinatorial analyses with some relaxations can easily establish an upper bound of $O(m^{k_1})$ for the number of dominated policies.
The algorithm to consider all dominated instances is not polynomial when the frequency $k_1$ of the manipulator in
the policy is not a constant.

We will use a dynamic programming technique to reduce the number of dominated instances to a polynomial without losing
an optimal solution. To do so, we need the following properties.


\section{Invariance Properties}
Our idea is a divide-and-conquer method.
We will partition the problem into two subproblems, the first part is to allocate the first $i$ items
and the second part is to allocate the remaining items.
We need to find the properties in the first part that keep the invariance of the second part.
Once we find these properties, we may only need to find the best allocation sequence of the first part for the manipulator satisfying these properties (for each fixed allocation sequence for the second part).
In this way, we may be able to use dynamic programming to reduce redundant cases  without losing an optimal solution.

It is easy to verify that the remaining problems are the same after executing two partial allocation sequences satisfying the following two conditions:

\begin{enumerate}
\item The number of items allocated to each agent (including the manipulator) is the same;
\item The set of items allocated to all the agents is the same.
\end{enumerate}

However, it is still hard to find all partial allocation sequences satisfying the above two conditions.
In order to get a polynomial-time algorithm, we add the third condition below

\begin{enumerate}
\item [3.] The last item allocated to each non-manipulator is the same.
\end{enumerate}
\begin{definition}[\textbf{Invariance Relation}]
Two (partial) allocation sequences are in the \emph{invariance relation} if they satisfy the above three conditions.
\end{definition}
 Recall that for an allocation sequence $\xi$, we use $\xi(i)$ to denote the partial allocation sequence of the first $i$ allocations.
\begin{lemma}\label{exchange}
Let $\xi$ be a feasible allocation sequence and  $\xi(i)$ ($1\leq i \leq m$) be a partial allocation sequence.
Let $\xi'(i)$ be another partial allocation sequence that is in the invariance relation with $\xi(i)$.
The allocation sequence $\xi'$ obtained by replacing $\xi(i)$ with $\xi'(i)$ in $\xi$ is still a feasible allocation sequence.
\end{lemma}

Since $\xi(i)$ and $\xi'(i)$ are in the invariance relation, we know that we will get the same remaining problem after executing them.
Thus we can exchange $\xi(i)$ and $\xi'(i)$ in larger allocation sequences.

The divide-and-conquer idea based on Lemma~\ref{exchange}
will be embedded in our dynamic programming algorithm. We will see that the algorithm will
only split the problem between segments.

\section{The Dynamic Programming Algorithm}
Equipped with the above properties, we are ready to describe the dynamic programming algorithm.
The main idea of the algorithm is still based on Lemma~\ref{search}. However, we will use Lemma~\ref{exchange} to reduce the number of
subproblems.

Recall that $m'$ is the number of  non-manipulator positions in the policy $\pi$.
For any integer $1\leq x \leq m'$, let $k(x)$ denote the times of the manipulator appearing during the first $x$ segments of the policy $\pi$, i.e., the period from the beginning of $\pi$ to the $x$th position of a non-manipulator. For any dominated instance $I'$, the occurrences of the manipulator in the first $x$ segments of the policy in $I'$
is at most $k(x)$.
Recall that $\pi^s(x)$ is the partial policy of the first $x$ segments of $\pi$.

We use $pro(x,y,i_2,\dots, i_n)$ to denote the set of all feasible partial allocation sequences satisfying the following conditions:
\begin{enumerate}
\item The core of the partial policy associated with the partial allocation sequence is the same as the core of $\pi^s(x)$;
\item The last allocation in the partial allocation sequence is to allocate an item to a non-manipulator;
\item Exactly $x$ items are allocated to non-manipulators and exactly $y$ items are allocated to the manipulator;
\item For each $j\in \{2,3,\dots, n\}$, the last item allocated to Agent $j$ is the $i_j$th item in its preference ranking, where $i_j$ can be 0 which means no item is allocated to Agent~$j$;
\item For each $r\in \{1,2,\dots, x\}$, during the first $r$ segments at most $k(r)$ items are allocated to the manipulator;
\item The partial allocation sequence is a greedy one.
\end{enumerate}

The domains of the parameters in $pro(x,y,i_2,\dots, i_n)$ are as follows: $x\in \{0, 1, \dots, m'\}$, $y \in \{0,1,\dots,k_1\}$
and $i_2, i_3,\dots,i_n\in \{0, 1, \dots, m\}$.
We may not describe the domains of the parameters when they are clear from the context.


Note that even all of $x$, $y$ and $i_j$ are fixed, the set $pro(x,y,i_2,\dots, i_n)$ may contain several different
allocation sequences, because the definition does not fix the positions of the $y$ manipulators in the corresponding (partial) policy.
We have the following property.
\begin{lemma}\label{invar}
Any two partial allocation sequences in $pro(x,y,i_2,\dots, i_n)$ are in the invariance relation.
\end{lemma}

Lemma~\ref{invar} can be proved by checking each of the three conditions of the invariance relation one by one, which is not hard and omitted here due to the limited space.

We use $opt(x,y,i_2,\dots, i_n)$ to denote a partial allocation sequence in $pro(x,y,i_2,\dots, i_n)$ where the manipulator gets
the best solution. Note that $pro(x,y,i_2,\dots, i_n)$ is possible to be empty and for this case we let $opt(x,y,i_2,\dots, i_n)=\perp$.

The allocation sequence $opt(x,y,i_2,\dots, i_n)$ even for $x=m'$ may not be a complete allocation sequence of length $m$, since $y$ may be smaller than $k_1$ and some allocations to the manipulator are still left.
In fact, $opt(x=m',y,i_2,\dots, i_n)$ is a partial allocation sequence only missing the last part of the allocations corresponding to the trivial segment of the policy.
We use $opt^*(x=m',y,i_2,\dots, i_n)$ to denote the complete allocation sequence obtained from
$opt(x=m',y,i_2,\dots, i_n)$ plus the $k_1-y$ allocations of the $k_1-y$ best remaining items to the manipulator.

The following two lemmas will say that the best allocation sequence among all $opt^*(x,y,i_2,\dots, i_n)$ with $x=m'$ will lead to the optimal solution
to the original instance.

\begin{lemma}\label{correct1}
For any $y \in \{0,1,\dots,k_1\}$ and $i_2, i_3,\dots,i_n\in \{0, 1, \dots, m\}$, if $opt(x=m',y,i_2,\dots, i_n)\neq \perp$, then $opt^*(x=m',y,i_2,\dots, i_n)$
is a greedy allocation sequence for an instance dominated by the original instance.
\end{lemma}
\pf{
By the definition, we know that $opt(x=m',y,i_2,\dots, i_n)$ is a greedy partial allocation sequence.
Since $opt^*(x=m',y,i_2,\dots, i_n)$ is obtained from $opt(x=m',y,i_2,\dots, i_n)$ by adding behind $k_1-y$ best allocations
to the manipulator, we know that $opt^*(x=m',y,i_2,\dots, i_n)$ is also greedy.
We consider the policy $\pi^*$ corresponding to the greedy allocation sequence $opt^*(x=m',k_1-z,i_2,\dots, i_n)$.
By the 5th item in the definition of $pro(x,y,i_2,\dots, i_n)$, we know that
for each $r\in \{1,2,\dots, x\}$, during the first $r$ segments at most $k(r)$ items are allocated to the manipulator.
This means $\pi^*$ is dominated by the original policy $\pi$.
}

\begin{lemma}\label{correct2}
 Let $I_c$ be a crucial instance corresponding to $I$, where the trivial segment in $I_c$
 consists of $z$ occurrences of the manipulator $(0\leq z \leq k_1)$. Assume that in the optimal solution to $I_c$, for each $j\in \{2,3,\dots, n\}$, the last item allocated to Agent $j$ is the $a_j$th item in its preference ranking.
 Then $opt^*(m',k_1-z,a_2,\dots, a_n)$ leads to the optimal solution to the original instance $I$.
\end{lemma}
\pf{
The greedy allocation sequence $S$ to $I_c$, also  leading to  an optimal solution to the original instance $I$, is a candidate for
$opt^*(m',y,a_2,\dots, a_n)$.
On the other hand, by Lemmas~\ref{correct1} and~\ref{lem1}, we know that  $opt^*(m',k_1-z,a_2,\dots, a_n)$ is not better than $S$. Since $opt^*(m',k_1-z,a_2,\dots, a_n)$ is chosen as the best one,
we know that $opt^*(m',k_1-z,a_2,\dots, a_n)$ is as good as $S$.
}

We can not directly compute an optimal solution to the original instance $I$ according to Lemma~\ref{correct2},
since we do not know the values of $a_j$ in Lemma~\ref{correct2}.
However, by Lemma~\ref{search}, Lemma~\ref{correct1} and Lemma~\ref{correct2}, we know that the best one among all $opt^*(x,y,i_2,\dots, i_n)$ with $x=m'$ will get an optimal solution to the original instance $I$.
So our algorithm contains the following three main steps.
\begin{enumerate}
\item [1] Compute all $opt(x,y,i_2,\dots, i_n)$ by calling the subalgorithm OPT;
\item [2] Compute all $opt^*(x,y,i_2,\dots, i_n)$ with $x=m'$ from $opt(m',y,i_2,\dots, i_n)$;
\item [3] Find the best one among all $opt^*(x,y,i_2,\dots, i_n)$ with $x=m'$.
\end{enumerate}

The subalgorithm OPT in Step 1 is a dynamic programming algorithm that
compute all $opt(x,y,i_2,\dots, i_n)$ in an order with a nonincreasing value of $x$.
Before presenting the whole procedure of OPT, we introduce the idea in the algorithm.

Assume that all $opt(x',y,i_2,\dots, i_n)$ for $x'< x$ have been computed.
We use the following idea to compute $opt(x,y,i_2,\dots, i_n)$.

We use $r$ to denote the non-manipulator on $x$th position of the core of $\pi$, i.e., the $x$th non-manipulator in $\pi$ is Agent~$r$. Assume that $opt(x,y,i_2,\dots, i_n)\neq\perp$. Let $\pi_x$ be the policy corresponding to $opt(x,y,i_2,\dots, i_n)$.
We further assume that the last segment of $\pi_x$ consists of $q$ occurrences of the manipulator and one occurrence of Agent~$r$.
Then $opt(x,y,i_2,\dots, i_n)$ is given by the allocations $L_1$ of items to the first $x-1$ segments of $\pi_x$ plus
the allocations $L_2$ of items to the last segment of $\pi_x$.

Since we require that the allocation sequence in
$opt(x,y,i_2,\dots, i_n)$ is greedy, we know that $L_2$ is given by $q$ allocations of the first $q$ remaining items on Agent~$r$'s preference ranking to the manipulator plus one allocation of the $(q+1)$th remaining item to Agent~$r$.
Furthermore, the last item allocated to Agent~$r$ must be the $i_r$th item in Agent~$r$'s preference ranking.
By Lemma~\ref{exchange}, Lemma~\ref{invar} and the fact that the utility function is additive, we know that $L_1$ is given by $opt(x-1,y-q,i_2,\dots,i^*_r,\dots, i_n)$ for some $i'_r\leq i_r-(q+1)$.

However, we do not know the value of $q$ and $i^*_r$. In the algorithm,
we try all possible values for $q$ and $i^*_r$. Lemma~\ref{search} can guarantee that the best one among them is the correct allocation sequence we are seeking for. The whole procedure of OPT is presented in Algorithm~\ref{alg}.

\begin{algorithm}[h]
    \caption{Subalgorithm OPT}\label{alg}
   \KwIn{An instance $I=(O, N, \pi, \{\succ_i\}_{i=1}^n)$ of \textsc{Best Response}.}
    \KwOut{To compute $opt(x,y,i_2,\dots, i_n)$ for all $x\in \{0, 1, \dots, m'\}$, $y \in \{0,1,\dots,k_1\}$ and $i_2, i_3,\dots,i_n\in \{0, 1, \dots, m\}$.}

\For {all values of $x,y,i_2,\dots,i_n$,}
         {$opt(x,y,i_2,\dots, i_n)\leftarrow\perp$;}

$opt(0,0,0,\dots, 0)\leftarrow \emptyset$, which is empty but feasible;

\For{$x=1$ to $m'$}
{
Let  Agent~$r$ be the non-manipulator on the $x$th position of the core of $\pi$;

\For{all $i_2, i_3,\dots,i_n\in \{0, 1, \dots, m\}$ and $0\leq y\leq k(x)$, }
  {
  \For { $q\in \{0,1,\dots, k(x)\}$}
      {
       \If {There is a value $i^*_r\leq i_r-(q+1)$ such that $opt(x-1,y-q,i_2,\dots,i^*_r,\dots, i_n)\neq \perp$, and after executing $opt(x-1,y-q,i_2,\dots,i^*_r,\dots, i_n)$, the $(q+1)$th remaining item on Agent~$r$'s preference ranking is  exactly the $i_r$th item on the whole preference ranking of Agent~$r$,}
         {Let $opt'$ be $opt(x-1,y-q,i_2,\dots,i^*_r,\dots, i_n)$ plus $q$ allocations of the first $q$ remaining items on Agent~$r$'s preference ranking to the manipulator and one allocation of the $(q+1)$th remaining item to Agent~$r$;

         Let $opt(x,y,i_2,\dots, i_n)$ be the best of $opt'$ and current $opt(x,y,i_2,\dots, i_n)$;
         }
      }

  }
}
\end{algorithm}

Next, we analyze the running time of the whole algorithm.
The algorithm contains three steps.

The first step is to compute $opt(x,y,i_2,\dots, i_n)$ for $i_2, i_3,\dots,i_n\in \{0, 1, \dots, m\}$, $x\in \{0, 1, \dots, m'=m-k_1\}$ and $y \in \{0,1,\dots,k_1\}$. In total, there are $(1+m)^{n-1}(m-k_1+1)(k_1+1)< (1+m)^{n+1}$ subproblems need to be solved.
For each subproblem $opt(x,y,i_2,\dots, i_n)$ with $x>0$, we compute them from Steps 3 to 9 in OPT.
In Step 6, we have $k(x)+1$ loops. For each loop, we may use at most $O(i_tm)$ time.
So for each subproblem $opt(x,y,i_2,\dots, i_n)$, our algorithm uses at most $O(m^3)$ time.
OPT runs in $O((1+m)^{n+4})$ time.

Step 2 takes at most $O((1+m)^{n+2})$ time to extend all $opt(x=m',y,i_2,\dots, i_n)$ to $opt^*(x=m',y,i_2,\dots, i_n)$.

Step 3 is to find the best one among all $opt^*(x=m',y,i_2,\dots, i_n)$, which can be done in  $O((1+m)^{n+1})$ time.

\begin{theorem}
\textsc{Best Response} can be solved in $O((1+m)^{n+4})$ time.
\end{theorem}

For each constant number $n$ of agents, \textsc{Best Response} is polynomially solvable.

\section{A $0.5$-Approximation Algorithm}
Although for each fixed number of agents, the manipulating sequential allocation problem can be solved in polynomial time, the running time is exponential in the number $n$ of agents.
When $n$ is large, the algorithm will still be slow. So we also consider approximation algorithm for the problem.
We prove that
\begin{theorem}
For any instance of \textsc{Best Response} with additive utility functions,
if the manipulator takes the truthful preference ranking as its picking strategy, it can get a bundle with the total utility being at least half of that of the optimal solution.
\end{theorem}
\pf{

Let $I'$ be the corresponding crucial instance of the input instance $I$.
By Corollary~\ref{cor1}, we know that an optimal solution to $I'$ is an optimal solution to $I$. By Lemma~\ref{lem1}, we also know
that a solution to $I$ is at least as good as that to $I'$ under the same picking strategy.
So we only need to prove the theorem holds for crucial instance $I'$ and next we assume that the input instance is crucial.

We use $\xi_A$ and $\xi_B$ to denote the allocations by taking the optimal picking strategy and by
taking the truthful preference as the picking strategy, respectively.
Let $A=\{a_1, a_2, \dots, a_{k_1}\}$ be the bundle obtained by $\xi_A$ and  $B=\{b_1, b_2, \dots, b_{k_1}\}$ be the bundle obtained by $\xi_B$, where we assume that the items in the above two sets are listed according to the picking order.

We first prove that for any index $1\leq i \leq k_1$ such that $a_{i} \succ_1  b_{i}$, the item $a_{i}$ is also in $B$.
Assume to the contrary that $a_{i}\not\in B$, which means that item $a_i$ is not taken into the solution in $\xi_B$.
The allocation of item $a_{i}$ in $\xi_A$ and the allocation of item $b_{i}$ in $\xi_B$ happen as the same position of the policy, say the $x$th position. In $\xi_B$, an item $b_{i}$ with $u(b_i)< u(a_i)$ is allocated at the $x$th position, which means
that $a_i$ has already been allocated to some agent before the $x$th position in $\xi_B$ as the picking strategy in $\xi_B$ is the truthful preference.
However, the instance is a crucial instance and the optimal allocation sequence in $\xi_A$ is greedy.
Item $a_i$ is impossible to be allocated to a non-manipulator before the $x$th position in $\xi_B$.
Then $a_i$ can only be allocated to the manipulator in $\xi_B$, which is a contradiction to the assumption
that $a_{i}\not\in B$. So the above claim holds.

Let $L=\{i_1, i_2, \dots, i_l\}$ be the set of indices $i_j$ such that $a_{i_j} \succ_1  b_{i_j}$.
Let $\overline{L}=\{1,2,\dots,k_1\}\setminus L$. Note that $\overline{L}$ is not empty and index 1 is always in $\overline{L}$.
By the above claim, we have that
$$\sum_{i\in L}u(a_i)< \sum_{b\in B}u(b).$$
By the definitions of $L$ and $\overline{L}$, we have that
$$\sum_{i\in \overline{L}}u(a_i)\leq  \sum_{i\in \overline{L}}u(b_i) \leq\sum_{b\in B}u(b).$$
By summing up the above two inequalities, we get that
$$\sum_{a\in A}u(a)< 2\sum_{b\in B}u(b).$$ 
}

We also give a simple example, where the approximation ratio cannot be $0.5+\epsilon'$ for any constant $\epsilon'>0$.
This will show the approximation ratio of $0.5$ is tight for the mechanism of using the truthful preference.
There are three items $O=\{g_1, g_2, g_3\}$ to be allocated to two agents $N=\{1,2\}$.
The preference rankings are $\succ_1: g_1, g_2, g_3$ and $\succ_2: g_2, g_3, g_1$.
The policy is $\pi: 121$.
The utility function of the manipulator is that $u(g_1)=1$, $u(g_2)=1-\epsilon$ and  $u(g_3)=\epsilon$.
If the manipulator use the picking strategy $g_2g_1g_3$, it can get items $g_2$ and $g_1$ with the utility $2-\epsilon$.
If the manipulator use the  truthful preference ranking $g_1g_2g_3$ as the picking strategy, it can only get items $g_1$ and $g_3$ with the utility $1+\epsilon$. The approximation ratio is ${\frac{1+\epsilon}{2-\epsilon}}=0.5+{\frac{1.5\epsilon}{2-\epsilon}}$,
where $\frac{1.5\epsilon}{2-\epsilon}$ can be arbitrarily small.

\section{Conclusion}
\textsc{Best Response} can be regarded as one of the hardest natural problems in manipulating sequential allocation problems,
since most other problems can be reduced to it.
It has been known for years that \textsc{Best Response} with only two agents can be solved in polynomial time. However, it took more effort to establish the NP-hardness of \textsc{Best Response} with an unbounded number of agents. In this paper, we complete the ``gap'' by showing that \textsc{Best Response} is polynomially solvable for any constant number of agents.
Furthermore, we show that we can always get a 0.5-approximation solution if taking the truthful preference ranking of the manipulator as its picking strategy. Furthermore, the ratio 0.5 is
tight as far as using the truthful preference ranking.
It may be interesting to consider the approximation ratio for
using the truthful response on more no-strategyproof problems.

%

\end{document}